\documentclass[aps,prb,twocolumn,psfig,showpacs,superscriptaddress]{revtex4}
\usepackage[UKenglish]{babel}
\usepackage{times}
\usepackage{graphicx}
\usepackage{float}
\usepackage{latexsym,amsmath,amssymb,bm,euscript}
\usepackage{color}
\usepackage{subfigure}
\usepackage{epstopdf}
\usepackage[colorlinks=true,linkcolor=blue,citecolor=blue]{hyperref}
\usepackage{hyperref}
\usepackage{soul}
\usepackage{ulem}

\begin{document}

\title{Emergent spin-1 trimerized valence bond crystal in the spin-1/2 Heisenberg model on the star lattice}

\author{Shi-Ju Ran}
\affiliation{ICFO-Institut de
	Ciencies Fotoniques, The Barcelona Institute of Science and
	Technology, 08860 Castelldefels (Barcelona), Spain}
\author{Wei Li}
\affiliation{Department of Physics, Key Laboratory of Micro-nano Measurement-Manipulation and Physics (Ministry of Education), and International Research Institute for Multidisciplinary Science, Beihang University, Beijing 100191, China}
\author{Shou-Shu Gong}
\affiliation{Department of Physics, Key Laboratory of Micro-nano Measurement-Manipulation and Physics (Ministry of Education), and International Research Institute for Multidisciplinary Science, Beihang University, Beijing 100191, China}
\affiliation{National High Magnetic Field Laboratory, Florida State University, Tallahassee, FL 32310}
\author{Andreas Weichselbaum}
\affiliation{Physics Department, Arnold Sommerfeld Center for Theoretical Physics, and Center for NanoScience, Ludwig-Maximilians-Universit, Munich 80333, Germany}
\author{Jan von Delft}
\affiliation{Physics Department, Arnold Sommerfeld Center for Theoretical Physics, and Center for NanoScience, Ludwig-Maximilians-Universit, Munich 80333, Germany}
\author{Gang Su}
\email[Corresponding author. ]{Email: gsu@ucas.ac.cn}
\affiliation{School of Physical Sciences, and CAS Center for Excellence in Topological Quantum Computation, University of Chinese Academy of Sciences, Beijing 100049, China}
\affiliation{Kavli Institute for Theoretical Sciences, University of Chinese Academy of Sciences, Beijing 100190, China}

\begin{abstract}
  We explore the frustrated spin-$\frac12$ Heisenberg model on the star
  lattice with antiferromagnetic (AF) couplings inside each triangle
  and ferromagnetic (FM) inter-triangle couplings ($J_e<0$), and
  calculate its magnetic and thermodynamic properties. We show that
  the FM couplings do not sabotage the magnetic disordering of the
  ground state due to the frustration from the AF interactions inside
  each triangle, but trigger a fully gapped
  inversion-symmetry-breaking trimerized valence bond crystal (TVBC)
  with emergent spin-1 degrees of freedom. We discover that with
  strengthening $J_e$, the system exhibits a universal scaling behavior either with or
  without a magnetic field $h$: the order parameter, the five critical
  fields that separate the $J_e$-$h$ ground-state phase diagram into
  six phases, and the excitation gap obtained by low-temperature
  specific heat, all depend exponentially on $J_e$. Our work implies that the spin-1 VBCs can be stabilized by introducing small FM couplings in the geometrically frustrated spin-$\frac12$ systems.
\end{abstract}

\maketitle

\section{Introduction} Two-dimensional (2D) spin-1/2 frustrated
magnetic systems are currently of great interest \cite{Balents2010,QMlect},
because they may realize exotic quantum states that do not
possess any semi-classical spin ordering \cite{Diep}, such as quantum
spin liquids (QSLs) or valence bond crystals (VBCs).  Leading
candidates for realizing such states are spin-$\frac12$ Heisenberg models
with competing interactions on, e.g.\ square, honeycomb and kagom\'{e}
lattices \cite{Nishimoto2013,Zhu2013,Gong2013,Gong2014_1,Kagome1,
  Kagome2, Kagome3, Kagome4,Kagome2017,Ran2007,U1QSL,Yasir2013,
  Yasir2014,HePRX2017,He2014,Gong2014,Bauer2014}. A particularly promising QSL
system that has been argued to have experimental realizations is the
kagom\'{e} Heisenberg antiferromagnet (KHAF) \cite{PRL_98_077204,
  PRL_98_107204, PRL_103_237201, PRB_82_144412, PRL_109_037208,
  Nature_492_7429, PRL_110_207208}. However the nature of its ground
state, i.e., a gapped Z$_2$ spin liquid \cite{Kagome1, Kagome2,
  Kagome3, Kagome4,Kagome2017} versus a gapless U(1) Dirac spin liquid
\cite{Ran2007, U1QSL, Yasir2013, Yasir2014,HePRX2017}, is still under debate.

\begin{figure}[tbp]
	\includegraphics[angle=0,width=1.0\linewidth]{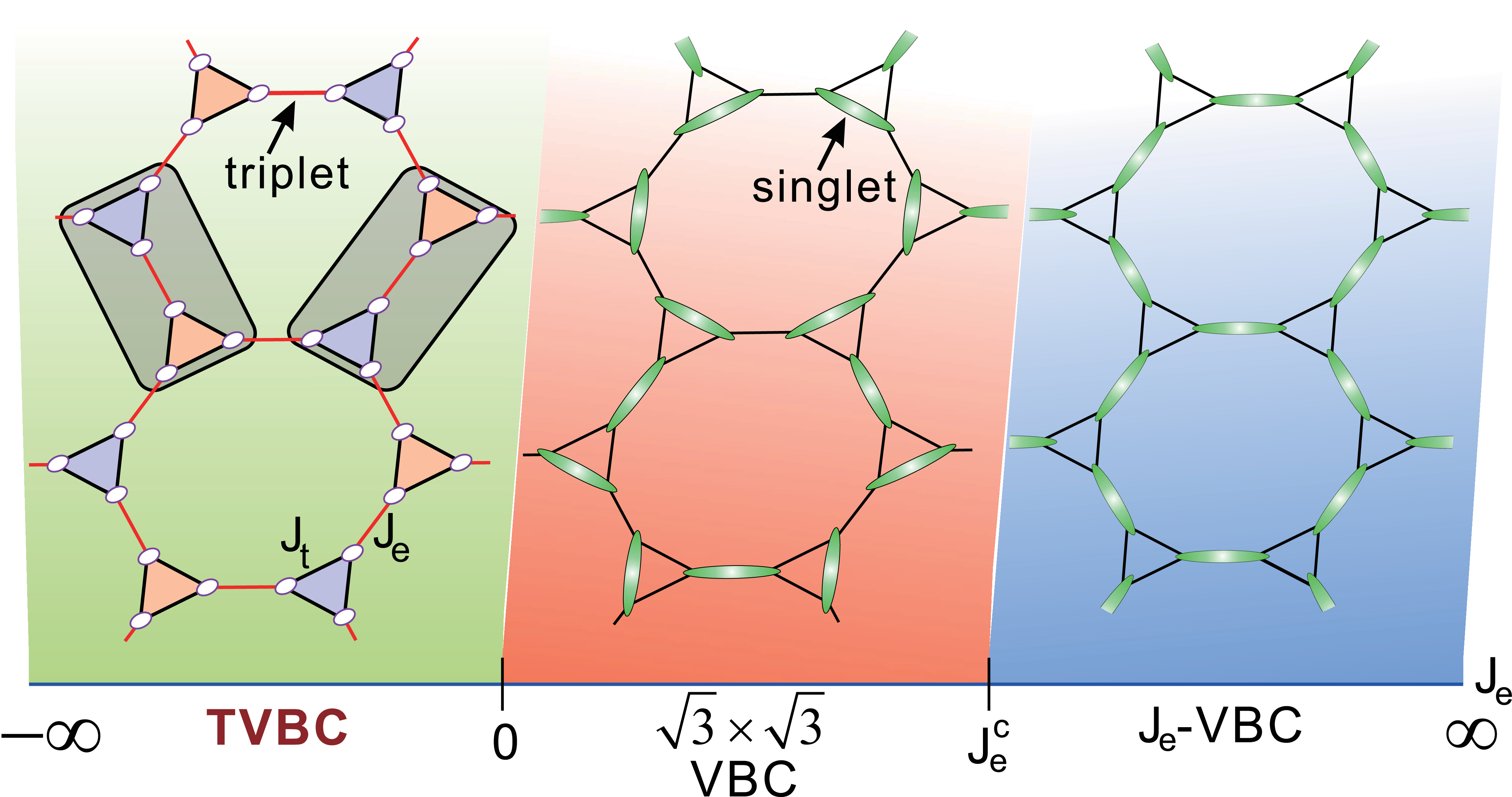}
	\caption{(Color online) The ground-state phase diagram of the star Heisenberg model. For $J_e>0$, previous studies show various possible VBCs and spin liquids, where one recent work found a $\sqrt{3} \times \sqrt{3}$ VBC and a $J_e$-bond VBC \cite{Yang2010,NoteStar}. The phase boundary $J_e^c$ has not been settled yet. For $J_e<0$, we show that the system is in a trimerized valence bond crystal (TVBC) phase, where a triplet appears at each $J_e$ bond and the inversion symmetry of up and down triangles (marked by blue and yellow, respectively) is broken.}
	\label{Phase}
\end{figure}

Another frustrated 2D quantum system of great potential interest is
the Heisenberg model on star lattice (Fig.~\ref{Phase}). The star lattice is an Archimedean lattice with all sites equivalent. It is also known as the (3-12) lattice, the Fisher lattice, the decorated hexagonal or expanded kagome lattice, and the triangle-honeycomb lattice, which are well summarized in Ref. [\onlinecite{star2008}]. Its physics
is arguably even richer than that of the KHAF, for several reasons: (a)
similar to the kagome lattice, the star lattice bears a high
geometrical frustration due to its triangle structure; (b) the star
lattice possesses a lower coordination number than the kagome lattice,
implying stronger fluctuations; (c) the star lattice naturally involves
two inequivalent bonds, which can lead to exotic quantum phases; (d)
various QSLs, such as the non-Abelian chiral spin liquid and the
double semion spin liquid, have been found in several models on a star
lattice, e.g., the Kitaev model and the quantum dimer model
\cite{star2008,Yao2012,Qi2014}; (e) a number of organic iron acetates have been synthesized in experiments \cite{Zheng2007}, which can be described by
the Heisenberg model on a star lattice.

However, the Heisenberg model on a star lattice has
not been fully explored yet. Recent research using the large-$N$
approximation and a Gutzwiller projected wave-function \cite{Yang2010}
only investigated the ground state for antiferromagnetic (AF)
inter-triangle couplings ($J_e > 0$), where a $J_e$-dimer VBC and a
$\sqrt{3} \times \sqrt{3}$ VBC phase \cite{NoteStar} were found
(Fig.~\ref{Phase}). However, the ground and thermal properties of the system for the ferromagnetic (FM) $J_e < 0$ are still unexplored. Recently, studying the effects of the FM couplings on 2D frustrated many-body systems has drawn lots of interests \cite{FMkagome,FMexp}. One of the issues we would like to address is whether the FM couplings can be adiabatically connect the spin-$1/2$ star model to the spin-$1$ kagom\'{e} model \cite{KagomeS12,KagomeS13,KagomeS14}.

The intrinsic importance of 2D frustrated many-body systems is matched
by the great technical challenges involved in studying them. One such
challenge is calculating thermodynamic properties, such as the specific heat and susceptibility. Most of the existing simulations of such systems are focused on the ground states. To compare with experiments, accurate simulations at finite temperature are strongly motivated, which are, however, scarce owing to the difficulties of such calculations \cite{ODTNS,NCD,FTTNS}.

In this work, we perform a comprehensive
study of the spin-$\frac12$ Heisenberg antiferromagnet on the star
lattice with FM inter-triangle couplings ($J_e<0$), calculating its
ground state and thermodynamic properties. We show that the FM
inter-triangle couplings do not sabotage the magnetic disordering of
the ground state that arises due to frustration generated by AF
intra-triangle couplings, but, remarkably, trigger a trimerized
valence bond crystal (TVBC) with emergent spin-1 degrees of freedom,
that breaks spatial inversion symmetry. We determine the phase diagram
of the system in a magnetic field and identify six phases. We uncover
a magnetization cusp on the boundary between the
inversion-symmetry-breaking and the non-inversion-symmetry-breaking
phases. We calculate the temperature dependence of the specific heat
and determine a non-magnetic gap by analyzing accurate results for the
low-temperature behavior of the specific heat. A scaling behavior versus $|J_e|$ is uncovered, evidenced by the large-$J_e$ dependence of a range of physical
quantities, such as the TVBC ``order parameter'', five critical fields
and the non-magnetic gap.

\section{Model and methods} The Hamiltonian of the star Heisenberg model reads
\begin{eqnarray}
 H=J_e\sum_{\langle ij\rangle \in J_e} S_i \cdot S_j + J_t\sum_{\langle lm\rangle \in J_t} S_l \cdot S_m - h \sum_{n} \hat{S}^z.
 \label{eq-H}
\end{eqnarray}
The first summation runs over all inter-triangle bonds, the
second over all intra-triangle bonds, and the third over all sites giving the magnetic field. 

Four different state-of-the-art algorithms are employed, including the SU(2) density matrix renormalization group (DMRG) \cite{DMRG} on a cylindrical geometry of finite size, the simple updates with and without non-Abelian SU(2) symmetry implemented \cite{PEPS,TRG2,Qspace} in the thermodynamic limit, and the network contractor dynamics (NCD) \cite{NCD} for the thermodynamics. These algorithms are designed for different purposes, and therefore operate differently. Overall consistency across these methods evidences a numerically unbiased, accurate, and comprehensive study.

\begin{figure}[tbp]
\includegraphics[angle=0,width=1.0\linewidth]{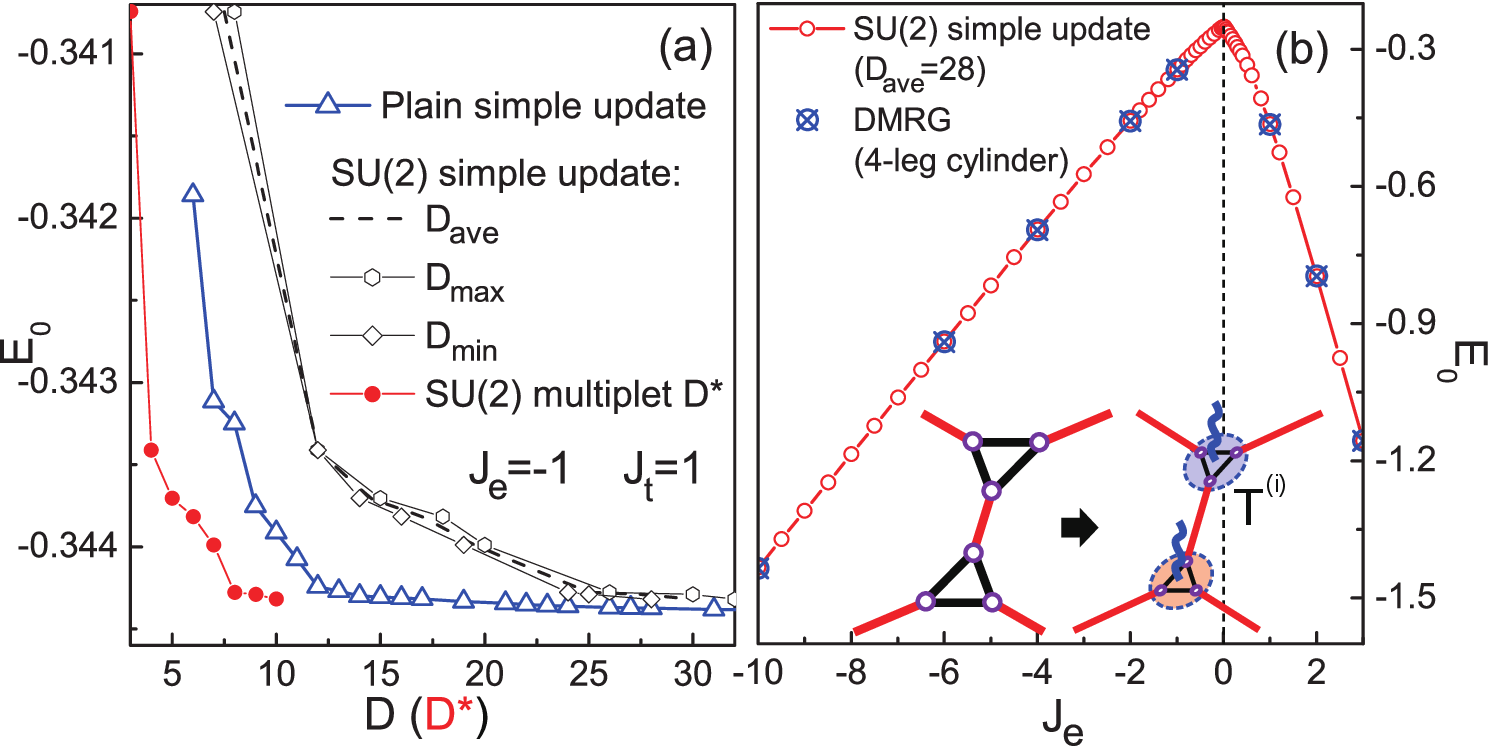}
\caption{(Color online) (a) The ground-state energy $E_0$ of the
star Heisenberg model for $J_e=-1$ and $J_t=1$ obtained by the
plain and SU(2) simple update algorithm, where $E_0$ converges
versus number of states ($D$) and number of multiplets ($D^{\ast}$
red solid dots), respectively, with clearly superior performance
of the SU(2)-based calculations. For comparison only, we also
translate the number of multiplets $D^{\ast}$ into the
corresponding actual number of states $D$ (black lines and
symbols). Taking a fixed $D^{\ast}$, $D$ of different virtual bonds may vary, according to the SU(2) fusion rules and the specific set of multiplets associated with each bond. We show the minimum, maximum and average value of
$D$ over the three virtual bond indices of a tensor. (b) $E_0$
versus $J_e$, obtained by SU(2) simple update and DMRG
simulations, which show very good agreement with each other in the
whole parameter range. The inset sketches the local tensors of the
TN state.}
\label{fig0}
\end{figure}

We employ tensor network (TN) \cite{TRG2} and DMRG \cite{DMRG} methods to simulate the ground state on the infinite lattice and cylindrical geometries, respectively. To be specific, the
TN representation of the ground state [inset of fig. \ref{fig0}(b)]
can be written as
\begin{eqnarray}
 | \psi \rangle = \sum_{\{s\}} {\rm{Tr}}_{\{a\}\in \mathrm{TN}} [\prod_{j} (T(j)^{s_{j,1}s_{j,2}s_{j,3}}_{a_{j,1}a_{j,2}a_{j,3}} | s_{j, 1}, s_{j, 2}, s_{j, 3}\rangle)],
 \label{eq-TNS}
\end{eqnarray}
where $T(j)$ is a ($d^3 \times D^3$) tensor residing on the $j$-th
triangle with physical dimension $d$ and ancillary bond dimension $D$,
containing all parameters of the TN state. The ancillary bonds
$\{a_{j,n}\}$ ($n=1,2,3$) carry the entanglement of the state and
$\rm{Tr}_{\{a\}\in \mathrm{TN}}$ denotes a contraction of all shared
$\{a_{j,n}\}$. The physical bonds $\{s_{j,n}\}$ $(n=1,2,3)$ represent
the three spins inside the $j$-th triangle with local basis
$|s_{j,n}\rangle$. Such a TN ansatz is called a projected
entangled-pair state (PEPS) \cite{PEPS}. The simple update algorithm \cite{TRG2}
provides an efficient way to optimize the PEPS by minimizing the
energy per site $E_0=\langle \psi|H|\psi \rangle$. The simple update has shown great efficiency and accuracy for simulating gapped systems. The observables such as magnetization can then be calculated with the PEPS.  

The method for finite-temperature simulations are implemented in a similar way. Each local tensor in the TN possesses two physical bonds that corresponds to the \textit{bra} and \textit{ket} space of the thermal state. We use the NCD approach \cite{NCD} to optimize the TN. The basic idea of NCD is to approximately encode the TN contraction problem into local self-consistent eigenvalue problems that can be efficiently solved. NCD shares a similar spirit with the simple update. Their mathematical background is the rank-1 decomposition that gives the optimal Bethe approximation of the corresponding TN \cite{TNrev}. The performance of such kind of approximation scheme is related to the speed of convergence to the fixed point when solving rank-1 decomposition, which is closely related to the value of gap. Thus, the algorithms show nice efficiency and accuracy for the gapped systems. The positions of the critical points that separate gapped phases can also be well determined due to the good performance within the gapped phases. At the critical points, however, it is still unclear how to optimize the tensors while keeping the criticality (such as the divergence of the correlation length) of the ansatz. Extracting critical information (e.g., critical exponents) is still a challenging task.


In addition, we implement SU(2) symmetry in TN states and related algorithms by using
QSpace techniques \cite{Qspace}: we impose SU(2) symmetry in every
single tensor index, retain the symmetry during imaginary time
evolutions and other tensor manipulations, and keep track of multiplets (instead of individual states) on the bonds. We only need to optimize the reduced tensors (instead of full tensors), andnthus reduce both the memory and CPU time dramatically \cite{Qspace, KagomeS12}.

\begin{figure*}[tbp]
	\includegraphics[angle=0,width=0.7\linewidth]{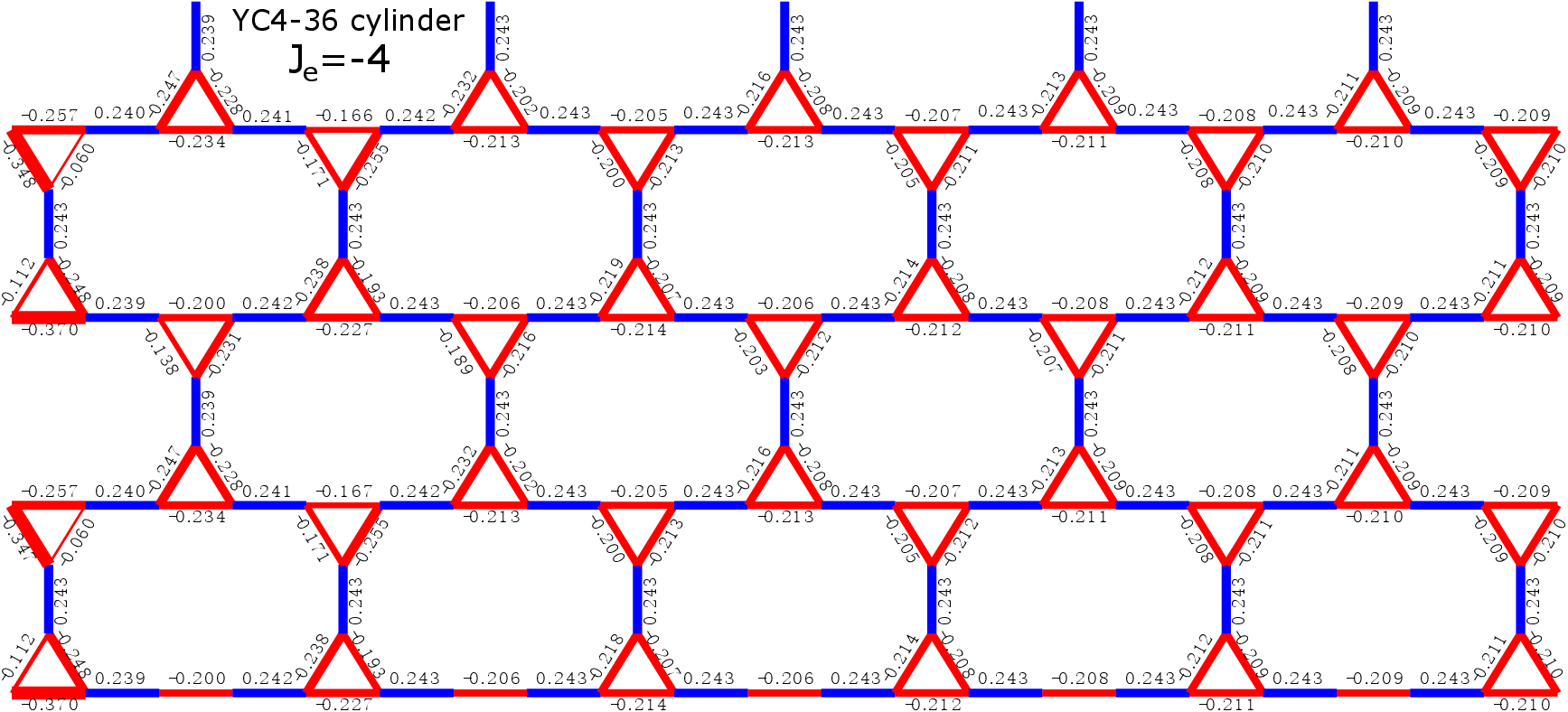}
	\caption{(Color online) The bond correlation $\langle S_i \cdot S_j \rangle$ versus distance $x$ from the pinned boundary (using $J_{\rm pin}=2$) on a YC4-36 cylinder for $J_e = -4$ calculated by DMRG keeping 2000 SU(2) multiplets. The thickness indicates the strength of the bond energies.}
	\label{fig2}
\end{figure*}

We compare the ground-state energy obtained by different
methods. In Fig.~\ref{fig0}(a), we show the energy obtained by plain
and SU(2) PEPS calculations, which both converge to the same
results. Note that for comparable number of states $D$, a lower ground state energy
can be obtained by plain PEPS as compared to SU(2) PEPS, since it
is allowed to break symmetries and hence has access to a larger
variational parameter space. However, the
results converge towards the same value for large $D$,
suggesting that as expected, the tensors eventually converge to tensors
that respect symmetries. This justifies the exploitation
of symmetries at significantly reduced overall numerical cost.

In Fig.~\ref{fig0}(b) we plot the energy obtained from SU(2) TN simulations and cylindrical DMRG for $-10.0\leq J_e \leq 3.0$,
which shows an excellent agreement in the whole region. Moreover, the appearance of a cusp in the energy curve at $J_e = 0$ indicates a
first-order phase transition.

\begin{figure}[tbp]
  \includegraphics[angle=0,width=1\linewidth]{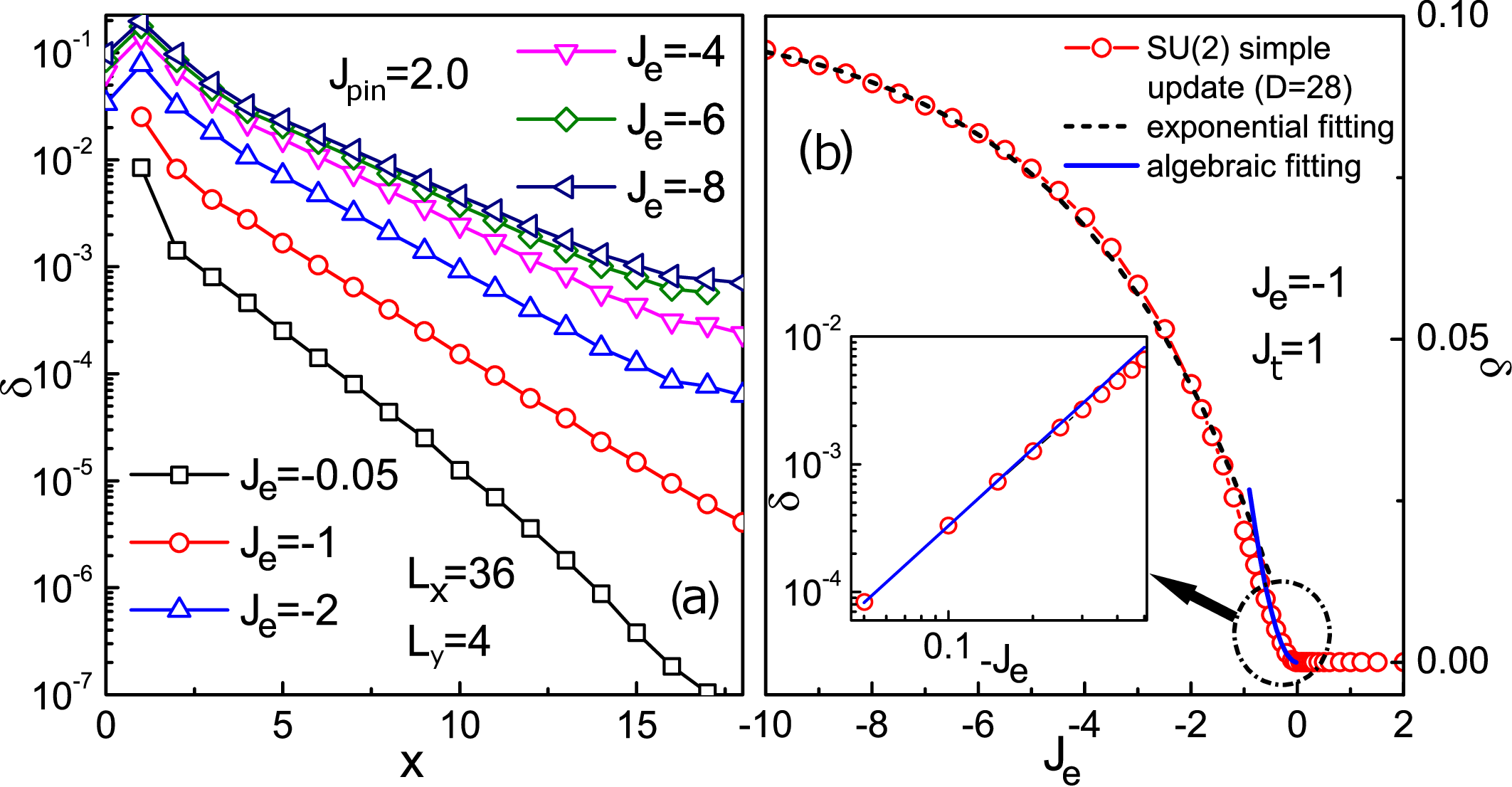}
  \caption{(Color online) (a) Log-linear plot of the inversion symmetry breaking parameter $\delta$ as function of the distance $x$ from the boundary for the YC4-36 cylinder with boundary pinning. (b) The $J_e$ dependence of $\delta$ obtained from SU(2) simple update simulations. As long as $J_e<0$, the system has a non-zero $\delta$. We find by fitting that for approximately $-J_e\gg 4$, $\delta$ fulfills the relation $\delta=\tilde{\delta}(1-e^{\mu J_e})$, where $\mu=0.28$ and $\tilde{\delta}=0.1$ that gives the value of $\delta$ for $J_e\rightarrow -\infty$. In contrast, for small $|J_e|$, we find $\delta = 0.03J_e^{2}$ for $J_e \rightarrow 0$ (see inset).} 
  \label{fig2b}
\end{figure}

\section{Spontaneous inversion symmetry breaking} We now study the
ground state of the star-lattice model, which is found to possess spontaneous
inversion symmetry breaking (SISB). It can be characterized by the
energy difference between the two kinds of triangles
$\delta \equiv |E^{\vartriangle} - E^{\triangledown}|$ where we have
$E^{\vartriangle (\triangledown)} = \langle \psi| \sum_{\langle ij
  \rangle \in \vartriangle (\triangledown)} H_{ij} | \psi \rangle$
per triangle with the summation running over all local interactions
$H_{ij}$ inside the up (down) triangles. We use DMRG to calculate the
cylinder system with the geometry shown in Fig.~\ref{fig2} (denoted
by YC4). To break the inversion symmetry between the up and down
triangles, we take the couplings inside the up triangles on the open
boundaries as $J_{\rm pin} = 2J_t$ ($J_t$ is the coupling constant for
all other triangles).

Then, we measure the decaying behavior of
$\delta$ from the boundary to the bulk. As shown in
Fig.~\ref{fig2b}(a), we find that $\delta$ decays quite slowly for large $-J_e$,
implying a large decay length. We checked that
different values of $J_{\rm pin}$ give the same decay length. Since the decay length for $\delta$ keeps increasing with increasing $-J_e$, our DMRG
calculations imply that the SISB of the ground state might survive on a wider or even infinite-size system. Based only on the DMRG results, however, it is difficult to determine whether the symmetry breaking persists in the thermodynamic limit. With decreasing $|J_e|$, $\delta$ decays faster. For small values of $|J_e|$, the SISB is too weak to identify on a small
cylinder. To provide more solid evidence, we thus employ the TN simulations on the infinite-size system.

The TN calculations, too, find a strong TVBC order for large $|J_e|$ [Fig.~\ref{fig2b}(b)], consistent with DMRG results. By fitting the order parameter $\delta$ with $-J_e\gg 0$, we find that $\delta$ fulfills an exponential behavior with $J_e$ as
\begin{eqnarray}
 \delta=\tilde{\delta}(1-e^{\mu J_e}),
 \label{eq-delta}
\end{eqnarray}
where we have $\mu=0.28$ and $\tilde{\delta}=0.1$. It indicates that the large $|J_e|$ couplings project each corresponding spin-$\frac12$ pair into an effective $S=1$ spin, and stabilize a TVBC. Interestingly, for the small $|J_e|$ region, the TN simulations show that the inversion symmetry is broken for any small $J_e < 0$, while such a symmetry is found to be intact for $J_e > 0$. To be specific, for $-J_e \to 0$, $\delta$ satisfies the algebraic relation $\delta=0.03J_e^{2}$, as shown in the inset of Fig. \ref{fig2b} (b). Our results not only support the TVBC ground state for the spin-$1$ kagom\'{e} model \cite{KagomeS12,KagomeS13,KagomeS14}, but also further show that such a TVBC is robust in the spin-$\frac12$ star model for any finite strength of the FM $J_e$ interactions. In other words, the TVBC survives with the fluctuations caused by the finiteness of $J_e$. In contrast to the spin-1 model, two spin-$\frac{1}{2}$'s in our system are not strictly projected into the spin-1's, especially for small $|J_e|$. 

Note that in the $J_e \to -\infty$ limit, each two spin-$\frac{1}{2}$'s connected by a ferromagnetic coupling are strictly mapped to the triplet states, i.e. a spin-1. Each antiferromagnetic coupling between two spin-$\frac{1}{2}$'s is then exactly mapped to the antiferromagnetic coupling of two spin-1's. The Hamiltonian of our star model becomes identical to that of the spin-$1$ antiferromagnetic Heisenberg model on kagome lattice.

\section{Ground-state phase diagram in magnetic fields}
\begin{figure}[tbp]
  \includegraphics[angle=0,width=1\linewidth]{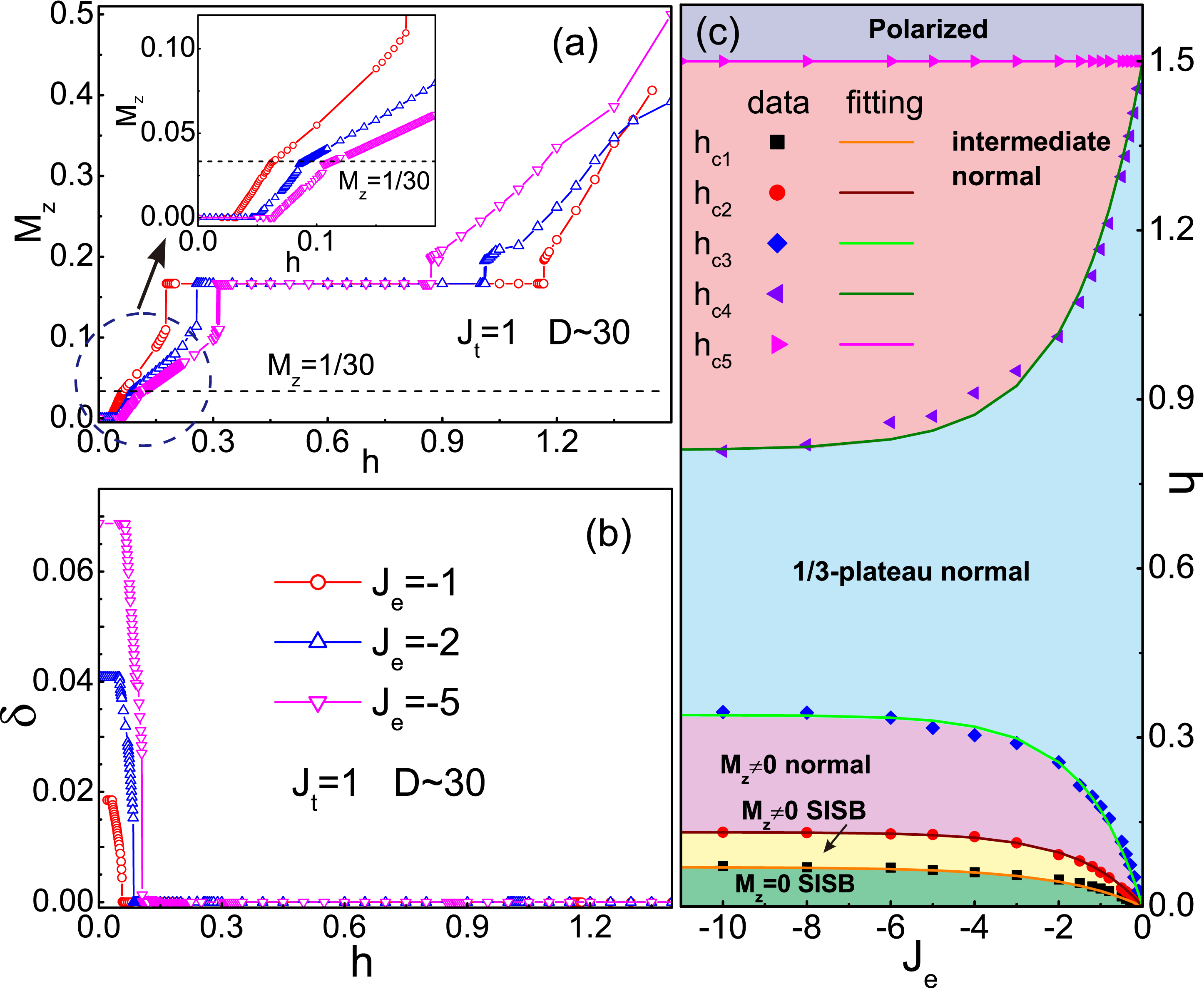}
  \caption{(Color online) The field-dependence of (a) the magnetization $M_z$ and (b) the TVBC order parameter $\delta$. Five critical fields $h_{ci}$ ($i=1,2,\cdots,5$) is determined by $M_z$ and $\delta$, which determine six phases in the $J_e$-$h$ diagram as shown in (c). For $0\leq h<h_{c1}$, $M_z=0$ and $\delta$ is intact. For $h_{c1}<h<h_{c2}$, $M_z$ increases, and $\delta$ starts to diminish and vanishes at $h = h_{c2}$, where the inversion symmetric and symmetry-breaking phases are separated and one always has $M_z=1/30$. 
  For $h_{c3}<h<h_{c4}$, the system is in a conventional $1/3$-plateau solid phase. Here, we use the simple update algorithm of PEPS with $D \sim 30$.}
\label{fig3}
\end{figure}

In a magnetic field, frustrated magnetic systems usually exhibit distinct features in the magnetization curve such as cusps \cite{Okunishi1999} and plateaus \cite{Oshikawa1997}, which reveal the exotic structure of the energy spectrum and distinguish different phases. We study the field dependence of the ground-state magnetization per site $M_z = \sum_{n} \langle \psi | \hat{S}^z_n |\psi \rangle/N$ and the energy difference $\delta$, as shown in Fig.~\ref{fig3}. Interestingly, we find a zero plateau corresponding to a finite spin gap, a cusp representing the restoration of inversion symmetry, and a $1/3$-plateau in the magnetization curve.

In the zero plateau region, $h < h_{c1}$, both $M_z$ and $\delta$
remain unchanged, indicating that there is a finite spin gap
protecting the TVBC state. With increasing $h$, the spin gap decreases
and eventually closes at $h = h_{c1}$.  For $h > h_{c1}$, $M_z$
becomes nonzero and $\delta$ starts to decrease. At $h = h_{c2}$, a
cusp appears in $M_z$ and $\delta$ vanishes, separating the SISB phase
from the $M_z \neq 0$ normal phases. A magnetization cusp has also been observed in
some one-dimensional frustrated magnetic systems having ground states
that break lattice symmetry, reflecting the novel energy dispersion of
the low-lying excitations \cite{Okunishi1999}. A first shoulder in the
magnetization occurs consistently around $M\simeq 1/30$. By further
increasing the field, we find a $1/3$-plateau corresponding to a
gapped solid state \cite{Oshikawa1997}. Based on the behaviors of
$M_z$ and $\delta$ we obtain the quantum phase diagram in the
$J_e$-$h$ plane, shown in Fig.~\ref{fig3}(c).

We find that the critical fields $h_{ci}$ ($i=1,2,\cdots,5$) also converge exponentially for large $|J_e|$,
\begin{eqnarray}
 h_{ci}=\tilde{h}_{ci}(1 - \alpha_i e^{\nu_i J_e}),
 \label{eq-hc}
\end{eqnarray}
as shown in Fig.~\ref{fig3}(c), with coefficients given in Table \ref{tab-Fields}. The scaling behavior of the critical fields strongly implies that the star Heisenberg model approaches the effective spin-1 model in an exponential manner, suggesting that the large $|J_e|$ represents a gapped system,
consistent with the existing works \cite{KagomeS12,KagomeS13,KagomeS14}.

\section{Specific heat} 
The calculation of specific heat is important for comparing with the experiments, where it can be directly measured by mature techniques, e.g. a thermal relaxation calorimeter. Thermal properties reflects not only the ground state, but also different physics at different temperature/energy scales.

\begin{figure}[tbp]
  \includegraphics[angle=0,width=1\linewidth]{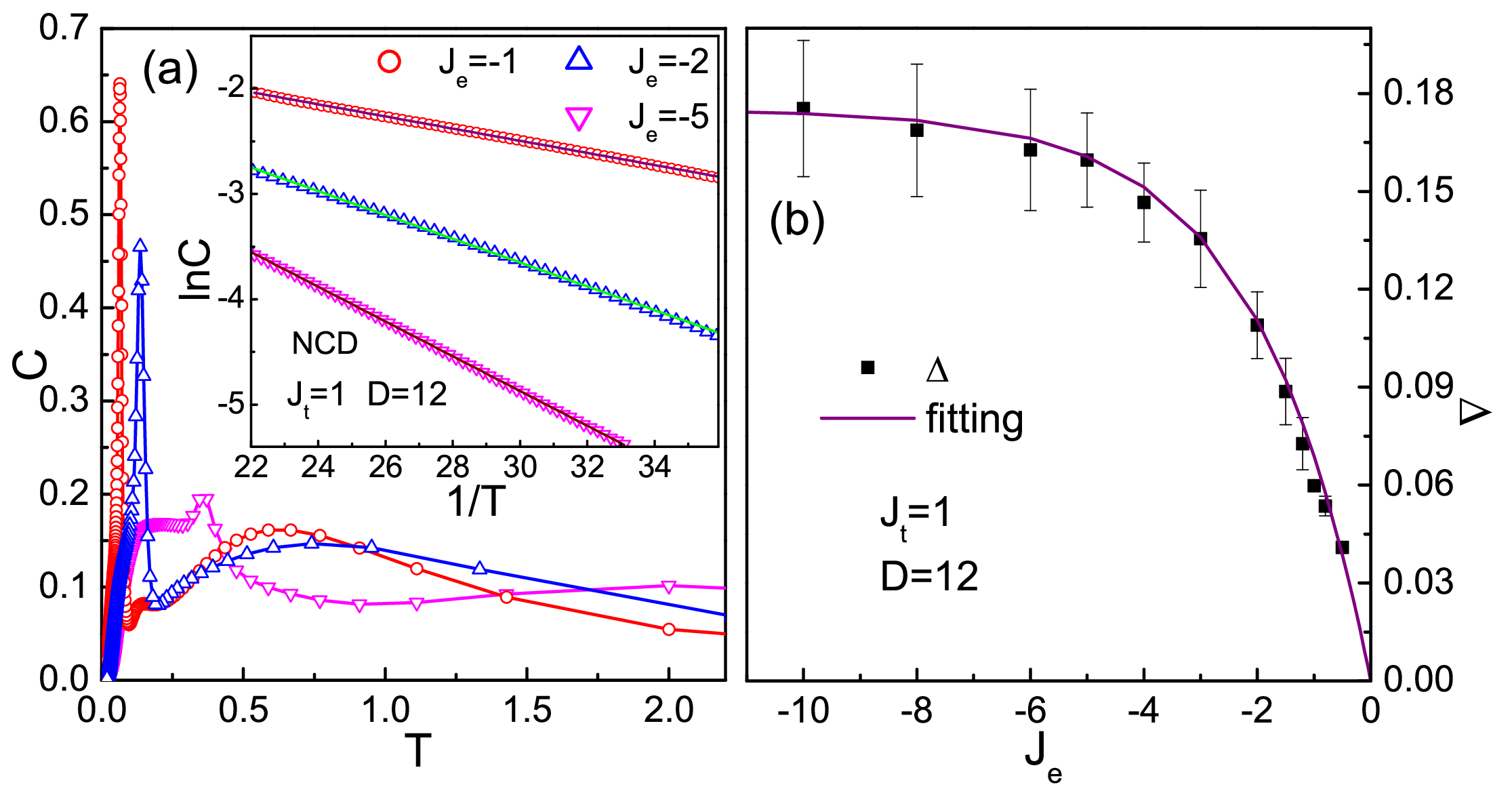}
  \caption{(Color online) (a) The temperature-dependence of specific heat $C$ for various $J_e$, where multi-peak structures are observed. We use the NCD algorithm with $D=12$. The position of the low-temperature peak moves to the higher temperature as $-J_e$ increases. Inset: the curve of $\ln C$ versus inverse temperature $1/T$. Below the low-temperature peak, one can see that the specific heat decays exponentially with $1/T$ as $\ln C = -\Delta/T + \rm{const}$, where $\Delta$ is the ($J_e$-dependent) excitation gap. (b) By fitting specific heat, the excitation gap $\Delta$ for different $J_e$ is obtained and shown to fulfill the relation $\Delta=\tilde{\Delta}(1-e^{\kappa J_e})$. The error bars are given by the linearity of $\ln C$ at the low temperatures \cite{ErrNote}. By fitting the $\Delta-J_e$ curve, we have $\tilde{\Delta}=0.175$ that gives the excitation gap in the $J_e \rightarrow -\infty$ limit and the constant $\kappa=0.5$.}
  \label{fig4}
\end{figure}

In Fig.~\ref{fig4}(a), we plot the calculated specific heat curves for various $J_e$. Changing $J_e$ from zero to $-\infty$, we observe that the low-temperature peak of the specific heat $C$ moves to higher temperature and merges with other peaks. From the inset, one can see that below the low-temperature peak, $\ln C$ depends linearly on the inverse
temperature $1/T$ as $\ln C = -\Delta/T + \rm{const.}$, indicating a
finite gap $\Delta$ that is consistent with the gapped TVBC ground
state. The $J_e$-dependence of $\Delta$ is given in
Fig.~\ref{fig4}(b). We observe again the exponential scaling behavior on $\Delta$ as
\begin{eqnarray}
 \Delta=\tilde{\Delta}(1-e^{\kappa J_e}),
 \label{eq-Delta}
\end{eqnarray}
where a fit yields $\kappa=0.5$ and $\tilde{\Delta}=0.17 \pm 0.02$
corresponding to the gap for $J_e \rightarrow -\infty$. Incidentally, simulations on the spin-1 kagome model also show a spin singlet gap $\tilde{\Delta}=0.1 \sim 0.2$
\cite{KagomeS1Private}. 

In principle, $\Delta$ is obtained from the low-temperature $C$, and should give the gap of the lowest excitation. In the given context, we expect $\Delta \sim h_{c1}$. Comparing with the critical fields, however, we find $\Delta \sim h_{c2}$, which should give the gap protecting spatial inversion symmetry of the up and down triangles. We provide the following scenario to explain our observations. The gaps for the excitations with $S \geq 1$ normally satisfy a linear relation as $\Delta(S) \sim S$, which leads to a non-zero magnetization when $h$ becomes larger than the spin gap. However, this linear relation may not be always true. Exceptions have been found when the system has ferromagnetic interactions \cite{FMex}. This means it is possible that non-zero magnetization appears at the $h$ smaller than the spin gap. In our model at $h=0$, the lowest excitation is the $S=1$ triplet state with the spin gap $\Delta \sim h_{c2}$ (this gap also protects the spatial inversion symmetry). Then non-zero magnetization appears with $h<h_{c2}$, say at $h_{c1}$ in our case, and the first excitation gap should be $h_{c2}$ at $h=0$.
		
On the other hand, we cannot completely exclude another possibility, where $\Delta \sim h_{c2}$ may be just a coincidence, caused by the computational error. We are unable to give a conclusive answer
here due to the lack precision of the existing methods.
Developing new approaches with higher accuracy for
2D models especially at the low temperatures would
be necessary. But in any case, we would like to stress
that this issue causes no harm to our main achievement,
which is the TVBC with an exponential scaling behavior.


\begin{table}[tbp]
  \caption{Values for the fitting parameters $\tilde{h}_{ci}$, $\alpha_i$ and $\nu_i$ of the critical fields [see Eq. (\ref{eq-hc})]. Note that as $h_{c5}=1.5$ is a constant, we have $\alpha_5=0$ and any $\nu_5$.}
  \begin{tabular*}{8cm}{@{\extracolsep{\fill}}lccccc}
  \hline\hline
    & $h_{c1}$ & $h_{c2}$ & $h_{c3}$ & $h_{c4}$ & $h_{c5}$ \\ \hline
  $\tilde{h}_{ci}$ & 0.07 & 0.132 & 0.34 & 0.81 & 1.5 \\
  $\alpha_i$ & 1 & 1 & 1 & -0.85 & 0\\
  $\nu_i$ & 0.5 & 0.65 & 0.7 & 0.6 & $\ast$
  \\ \hline\hline
  \label{tab-Fields}
  \end{tabular*}
\end{table}

\section{Conclusions} In this work, we discover an emergent spin-1 TVBC with spontaneous lattice inversion symmetry breaking in the spin-$\frac12$ star Heisenberg model with FM inter-triangle couplings, and study its ground-state and thermodynamic properties. We employ four different algorithms including SU(2) DMRG, simple update of the TN state with and without SU(2) symmetry, and NCD. Rich properties that define the exotic TVBC phase are revealed, including fruitful phases in a magnetic field, the magnetic cusps at $M_z \simeq 1/30$ and the universal exponential scaling behavior. Our work implies that spin-1 VBCs can be stabilized in the geometrically frustrated spin-$\frac12$ star-lattice systems with an arbitrary strength of the FM interactions. Moreover, our calculations of the specific heat provide useful data at finite temperatures which can be compared directly with the future experiments.


\section*{Acknowledgments} We are indebted to C. Peng for useful discussions. This work was supported in part by the MOST of China (Grants No. 2012CB932900 and No. 2013CB933401), the Strategic Priority Research Program of the Chinese Academy of Sciences (Grant No. XDB07010100), and NSFC Grant No. 11474249. S.S.G. was supported by the National Science Foundation through grant DMR-1408560, and the National High Magnetic Field Laboratory that is supported by NSF DMR-1157490 and the State of Florida. W.L. was supported by SFB-TR12 and the National Natural Science Foundation of China (Grant No. 11504014), A.W. by DFG WE4819/1-1 and WE4819/1-2. S.J.R. was supported by ERC AdG OSYRIS (ERC-2013-AdG Grant No. 339106), the Spanish MINECO grants FOQUS (FIS2013-46768-P), FISICATEAMO (FIS2016-79508-P), and ``Severo Ochoa'' Programme (SEV-2015-0522), Catalan AGAUR SGR 874, Fundaci\'o Cellex, EU FETPRO QUIC, EQuaM (FP7/2007-2013 Grant No. 323714), CERCA Programme / Generalitat de Catalunya, and Fundaci\'o Catalunya - La Pedrera $\cdot$ Ignacio Cirac Program Chair.


\begin{thebibliography}{99}

\bibitem{Balents2010} L. Balents, Nature \textbf{464}, 199 (2010).

\bibitem{QMlect} J. Richter, J. Schulenburg and A. Honecker, Lect.\ Notes Phys.\ {\bf 645}, 85-153 (2004).

\bibitem{Diep} H. T. Diep,
\textit{Frustrated spin systems}. World Scientific, 2004.





\bibitem{Kagome1} S. Yan, D. Huse, and S. R. White,
Science \textbf{332}, 1173-1176 (2011).

\bibitem{Kagome2} S. Depenbrock, I. P. McCulloch, and U. Schollw\"{o}ck, Phys. Rev. Lett. \textbf{109}, 067201 (2012).

\bibitem{Kagome3} H. C. Jiang, Z. H. Wang and L. Balents,
Nat. Phys. \textbf{8}, 902-905 (2012).

\bibitem{Kagome4} S. Nishimoto, N. Shibata and C. Hotta,
Nat. Commun. \textbf{4}, 2287 (2013).

\bibitem{Kagome2017} J. W. Mei, J. Y. Chen, H. He and X. G. Wen, Phys. rev. B \textbf{95},
235107 (2017).

\bibitem{Ran2007} Y. Ran, M. Hermele, P. A. Lee, and X. G. Wen,
Phys. Rev. Lett. {\bf 98}, 117205 (2007).

\bibitem{U1QSL} M. Hermele, Y. Ran, P. A. Lee, and X.-G. Wen, Phys. Rev. B \textbf{77}, 224413 (2008).

\bibitem{Yasir2013} Y. Iqbal, F. Becca, S. Sorella, and D. Poilblanc,
Phys. Rev. B {\bf 87}, 060405(R) (2013).

\bibitem{Yasir2014} Y. Iqbal, D. Poilblanc, and F. Becca,
Phys. Rev. B {\bf 89}, 020407(R) (2014).

\bibitem{HePRX2017} Y. C. He, M. P. Zaletel, M. Oshikawa, and F. Pollmann, Phys. Rev. X \textbf{7}, 031020 (2017).

\bibitem{He2014} Y. C. He, D. N. Sheng, and Y. Chen,
Phys. Rev. Lett. {\bf 112}, 137202 (2014).

\bibitem{Gong2014} S. S. Gong, W. Zhu, and D. N. Sheng,
Sci. Rep. {\bf 4}, 6317 (2014).

\bibitem{Bauer2014} B. Bauer, L. Cincio, B. P. Keller, M. Dolfi, G. Vidal, S. Trebst, and A. W. W. Ludwig,
Nat. Commu. {\bf 5}, 5137 (2014).

\bibitem{Nishimoto2013} R. Ganesh, Jeroen van den Brink, and S. Nishimoto,
Phys. Rev. Lett. \textbf{110}, 127203 (2013).

\bibitem{Zhu2013} Z. Y. Zhu, D. A. Huse, and S. R. White,
Phys. Rev. Lett. \textbf{110}, 127205 (2013).

\bibitem{Gong2013} S. S. Gong, D. N. Sheng, O. I. Motrunich, and M. P. A. Fisher,
Phys. Rev. B \textbf{88}, 165138 (2013).

\bibitem{Gong2014_1} S. S. Gong, W. Zhu, D. N. Sheng, O. I. Motrunich, and M. P. A. Fisher,
Phys. Rev. Lett. \textbf{113}, 027201 (2014).


\bibitem{PRL_98_077204} P. Mendels, F. Bert, M. A. de Vries, A. Olariu, A. Harrison, F. Duc, J. C. Trombe, J. S. Lord, A. Amato, and C. Baines,
Phys. Rev. Lett. \textbf{98}, 077204 (2007).

\bibitem{PRL_98_107204} J. S. Helton, K. Matan, M. P. Shores, E. A. Nytko, B. M. Bartlett, Y. Yoshida, Y. Takano, A. Suslov, Y. Qiu, J.-H. Chung,
D. G. Nocera, and Y. S. Lee,
Phys. Rev. Lett. \textbf{98}, 107204 (2007).

\bibitem{PRL_103_237201} M. A. de Vries, J. R. Stewart, P. P. Deen, J. Piatek, G. N. Nilsen, H. M. Ronnow, and A. Harrison,
Phys. Rev. Lett. \textbf{103}, 237201 (2009).

\bibitem{PRB_82_144412} D. Wulferding, P. Lemmens, P. Scheib, J. R\"{o}der, P. Mendels, S. Chu, T. Han, and Y. S. Lee,
Phys. Rev. B \textbf{82}, 144412 (2010).

\bibitem{PRL_109_037208} B. F\r{a}k, E. Kermarrec, L. Messio, B. Bernu, C. Lhuillier, F. Bert, P. Mendels,
B. Koteswararao, F. Bouquet, J. Ollivier, A. D. Hillier, A. Amato, R. H. Colman, and A. S. Wills,
Phys. Rev. Lett. \textbf{109}, 037208 (2012).


\bibitem{Nature_492_7429} T. H. Han, J. S. Helton, S. Chu, D. G. Nocera, J. A. Rodriguez-Rivera, C. Broholm, and Y. S. Lee, Nature \textbf{492}, 7429 (2012).

\bibitem{PRL_110_207208} L. Clark, J. C. Orain, F. Bert, M. A. De Vries, F. H. Aidoudi, R. E. Morris, P. Lightfoot, J. S. Lord, M. T. F. Telling, P. Bonville, J. P. Attfield, P. Mendels, and A. Harrison, Phys. Rev. Lett. \textbf{110}, 207208 (2013).


\bibitem{star2008} John Ove Fjaerestad, Arxiv:0811.3789.

\bibitem{Yao2012} H. Yao and S. A. Kivelson, Phys. Rev. Lett. {\bf 108}, 247206 (2012).

\bibitem{Qi2014} Y. Qi, Z. C. Gu, and H. Yao, Arxiv:1406.6364.

\bibitem{Zheng2007} Y. Z. Zheng, M. L. Tong, W. Xue, W. X. Zhang, X. M. Chen, F. Grandjean, and G. J. Long, Angew. Chem. Int. Ed. {\bf 46}, 6076 (2007).

\bibitem{Yang2010} B. J. Yang, A. Paramekanti, and Y. B. Kim,
Phys. Rev. B {\bf 81}, 134418 (2010).

\bibitem{NoteStar} Note that some former researches by several methods also indicated other possible VBCs and QSLs. See e.g. J. Richter, J. Schulenburg, A. Honecker, and D. Schmalfu{\ss} Phys. Rev. B {\bf 70}, 174454 (2004); G. Misguich and P. Sindzingre, J. Phys.: Condens. Matter {\bf 19}, 145202 (2007); T. P. Choy and Y. B. Kim, Phys. Rev. B {\bf 80}, 064404 (2009).

\bibitem{FMkagome} S. Bieri, L. Messio, B. Bernu, and C. Lhuillier,
Phys. Rev. B \textbf{92}, 060407(R) (2015).

\bibitem{FMexp} D. Boldrin, B. F{\aa}k, M. Enderle, S. Bieri, J. Ollivier, S. Rols, P. Manuel, and A. S. Wills, Phys. Rev. B \textbf{91}, 220408(R) (2015).

\bibitem{KagomeS12} T. Liu, W. Li, A. Weichselbaum, J. von Delft, G. Su, Phys. Rev. B \textbf{91}, 060403(R) (2015).

\bibitem{KagomeS13} H. J. Changlani and A. M. L\"{a}uchli, Phys. Rev. B \textbf{91}, 100407(R) (2015).

\bibitem{KagomeS14} T. Picot and D. Poilblanc, Phys. Rev. B \textbf{91}, 064415 (2015).

\bibitem{ODTNS} S. J. Ran, W. Li, B. Xi, Z. Zhang, and G. Su, Phys. Rev. B \textbf{86}, 134429 (2012).

\bibitem{NCD} S. J. Ran, B. Xi, T. Liu, and G. Su, Phys. Rev. B \textbf{88}, 064407 (2013).

\bibitem{FTTNS} P. Czarnik, L. Cincio, and J. Dziarmaga, Phys. Rev. B \textbf{86}, 245101 (2012); P. Czarnik and J. Dziarmaga, Phys. Rev. B \textbf{90}, 035144 (2014); \textbf{92}, 035120 (2015).

\bibitem{DMRG} S. R. White, Phys. Rev. Lett. \textbf{69}, 2863 (1992), Phys. Rev. B. \textbf{48}, 10345 (1993).

\bibitem{PEPS} F. Verstraete and J. I. Cirac, arXiv:cond-mat/0407066; J. Jordan, R. Or\'{u}s, G. Vidal, F. Verstraete, and J. I. Cirac, Phys. Rev. Lett. \textbf{101}, 250602 (2008).

\bibitem{Qspace} A. Weichselbaum, Anna. of Phys. \textbf{327}, 2972-3047 (2012).


\bibitem{TRG2} H. C. Jiang, Z. Y. Weng, and T. Xiang, Phys. Rev. Lett. \textbf{101}, 090603 (2008); Z. Y. Xie, H. C. Jiang, Q. N. Chen, Z. Y. Weng, and T. Xiang, \textit{ibid} \textbf{103}, 160601 (2009).

\bibitem{TNrev} S. J. Ran, E. Tirrito, C. Peng, X. Chen, G. Su, and M. Lewenstein, arXiv:1708.09213.


\bibitem{Okunishi1999} K. Okunishi, Y. Hieida, and Y. Akutsu,
Phys. Rev. B {\bf 60}, R6953(R) (1999).

\bibitem{Oshikawa1997} M. Oshikawa, M. Yamanaka, and I. Affleck,
Phys. Rev. Lett. {\bf 78}, 1984 (1997).

\bibitem{KagomeS1Private} Private communications with H. J. Changlani and A. M. L\"{a}uchli.

\bibitem{ErrNote} Note that the error bars only indicate
the error from 
fitting. The total error should also 
include the error inherent to the choice of the algorithm
and, due to finite numerical resources,
the Trotter-Suzuki error, and the accumulated truncation error.
There is no known method to estimate 
their combined effect on the total error.

\bibitem{FMex} N. Shannon, T. Momoi, and P. Sindzingre, Phys. Rev. Lett. {\bf 96}, 027213 (2006).

\end{thebibliography}
\end{document}